\documentclass[aps,pra,reprint,superscriptaddress,longbibliography]{revtex4-1}

\usepackage[colorlinks=true, allcolors=blue]{hyperref} 
\hypersetup{pdfauthor={Walasik el al.},pdftitle={Continuous supersymmetric transformation for photonic design}}

\usepackage{wasysym}
\usepackage{amsfonts}

\usepackage{cleveref}
\usepackage[export]{adjustbox}

\usepackage{color}

\crefformat{equation}{Eq.~(#2#1#3)}
\crefformat{figure}{Fig.~#2#1#3}

\Crefformat{equation}{Equation~(#2#1#3)}
\Crefformat{figure}{Figure~#2#1#3}

\begin{document}


\title{Continuous supersymmetric transformation for photonic design}


\author{Wiktor Walasik}
\email[]{wiktor.walasik@duke.edu}
\affiliation{Department of Electrical and Computer Engineering, Duke University, Durham, North Carolina 27708, USA}

\author{Nitish Chandra}
\affiliation{Department of Electrical and Computer Engineering, Duke University, Durham, North Carolina 27708, USA}

\author{Bikashkali Midya}
\affiliation{Department of Materials Science and Engineering, University of Pennsylvania, Philadelphia,
	PA 19104, USA}

\author{Liang Feng}
\affiliation{Department of Materials Science and Engineering, University of Pennsylvania, Philadelphia,
	PA 19104, USA}

\author{Natalia M. Litchinitser}
\email[]{natalia.litchinitser@duke.edu}
\affiliation{Department of Electrical and Computer Engineering, Duke University, Durham, North Carolina 27708, USA}

\date{\today}

\begin{abstract}
We propose to use a continuous supersymmetric (SUSY) transformation of a dielectric permittivity profile in order to design a photonic mode sorter. 
The iso-spectrality of the SUSY transformation ensures that modes of the waveguide preserve their propagation constants while being spatially separated. 
This global matching of the propagation constants, in conjunction with the adiabatic modification of the refractive index landscape along the propagation direction results in the negligible modal cross-talk and low scattering losses in the sorter. 
We show that a properly optimized SUSY mode sorter outperforms a standard asymmetric Y-splitter by reducing the cross-talk by at least two orders of magnitude. 
Moreover, the SUSY sorter is capable of sorting either transverse-electric or transverse-magnetic polarized modes and operates in a broad range of wavelengths.
The design proposed here paves the way toward efficient signal manipulation in integrated photonic devices.
\end{abstract}


\maketitle


\section{Introduction}

The rapidly growing demand for high-capacity optical-transmission technologies~\cite{Agrell16} sparked the growth of integrated~\cite{Lifante03} and silicon~\cite{Reed04} photonics. 
Efficient on-chip manipulation of optical signals requires development of high fidelity Y-junctions~\cite{Burns74,Love12,Riesen:12,Driscoll:13,Zhang:13}, photonic lanterns~\cite{Birks:15}, mode filters and multiplexers~\cite{Dai17}, and interferometers~\cite{Fischer94} that are optimized in terms of device length and efficiency~\cite{Sun:09,Martinez-Garaot:14,Martinez-Garaot:17,Chung:17}.
Recently, integrated mode converters~\cite{Heinrich:14a,Queralto:17}, filters, and beam splitters~\cite{Queralto:18} based on the principles of supersymmetry (SUSY) were proposed.

The notion of SUSY emerged first in the quantum field theory where it related the bosonic and fermionic degrees of freedom~\cite{Weinberg05}. 
Later, SUSY was used as an analytical tool in quantum mechanics~\cite{Cooper01}, allowing for discovery of new families of reflectionless~\cite{MAYDANYUK05} and periodic potentials~\cite{Dunne98a,Khare04}.
The similarities between the Schr\"odinger equation and the Helmholtz equation describing electromagnetic waves enabled application of SUSY in optical systems~\cite{CHUMAKOV94,Miri13,Miri:14,Laba14}. 
Various problems in photonics were addressed using SUSY, including creation of topologically protected midgap states~\cite{Midya:18} and optimization of semiconductor quantum-well cascade lasers~\cite{Milanovic96,Tomic97,Bai:08}, laser arrays~\cite{El-Ganainy15,Teimourpour16,Hokmabadi2:18} and other coupled systems~\cite{Walasik:18}.

To date, the majority of applications employ the unbroken SUSY that relates partners supporting the same set of eigenstates with the exception of the fundamental state. 
Whereas, the broken SUSY allows one to generate  families of exactly iso-spectral potentials described by a continuous parameter. 
Optical applications of the broken SUSY remain mostly unexplored, with the exception of the scattering structures studied in Ref.~\cite{Miri:14}. 

Here, we propose a design of a mode sorter based on the continuous SUSY transformation in the broken regime. 
We introduce the permittivity distribution that changes adiabatically~\cite{Menchon16} along the propagation direction to minimize the scattering losses and cross-talk and that is described by a continuous SUSY transformation of the initial permittivity profile. 
The latter ensures that the propagation constants of the modes to be sorted are preserved along the length of the sorter.
We demonstrate that, as a result of this global matching of the propagation constants, the SUSY design allows for reduction of the modal cross-talk by two orders of magnitude compared with a standard asymmetric Y-splitter~\cite{Burns74}.
Moreover, the SUSY mode sorter operates for both transverse-electric (TE) and transverse-magnetic (TM) light polarization, and it shows low losses and modal cross-talk over a broad wavelength range.
Finally, we show that sorting of multiple modes~\cite{Love12} is possible with the SUSY based design. 
Compared with the previous SUSY based modes sorters~\cite{Queralto:17,Queralto:18}, our design offers similar performance with an order of magnitude smaller sorter length and can separate modes without losing energy via radiative modes.
It is important to stress that SUSY-based design method presented here enables the cross-talk reduction by a systematic choice of the transverse waveguide profiles at different cross-sections along the length of the sorter ensuring that the propagation constants of the modes are preserved. 
On the contrary, previously used optimization approaches mostly use a fixed rectangular waveguide profiles and rely on optimizing the spacing or coupling constants between the waveguides, while the input and ouput modes have different propagation constants~\cite{Sun:09,Martinez-Garaot:17,Chung:17}.
Our SUSY-based design might find application in splitting and recombining different modes of integrated devices that require matching of the propagation constants.

\section{Mode sorter design}

Let us consider a planar waveguide (invariant along the $y$-direction) defined by the permittivity distribution~$\epsilon_{\infty}(x)$. 
For the TE polarization of light, the modes of such a waveguide are fully defined by $E_y(x,z) = \phi(x)\exp(i\beta z)$, where the propagation constant $\beta$ is related to the effective index of the mode as $\beta =k_0 n_{\rm{eff}}$ and the transverse electric field profile is denoted by $\phi(x)$. 
Here, $k_0 = 2\pi/\lambda$ and $\lambda$ denotes the free-space wavelength of light.
The mode profiles $\phi$ can be found using the Helmholz equation
\begin{equation}
[\partial_x^2 + k_0^2\epsilon_{\infty}(x)]\phi_j = \beta_j^2 \phi_j,
\label{eqn:Helm}
\end{equation}
where the index $j$ enumerates the modes.
Unless stated otherwise, the results presented in this work are obtained for TE polarization.

\begin{figure}[!t]
	\includegraphics[width = \columnwidth, clip = true, trim = {10 160 10 20}]{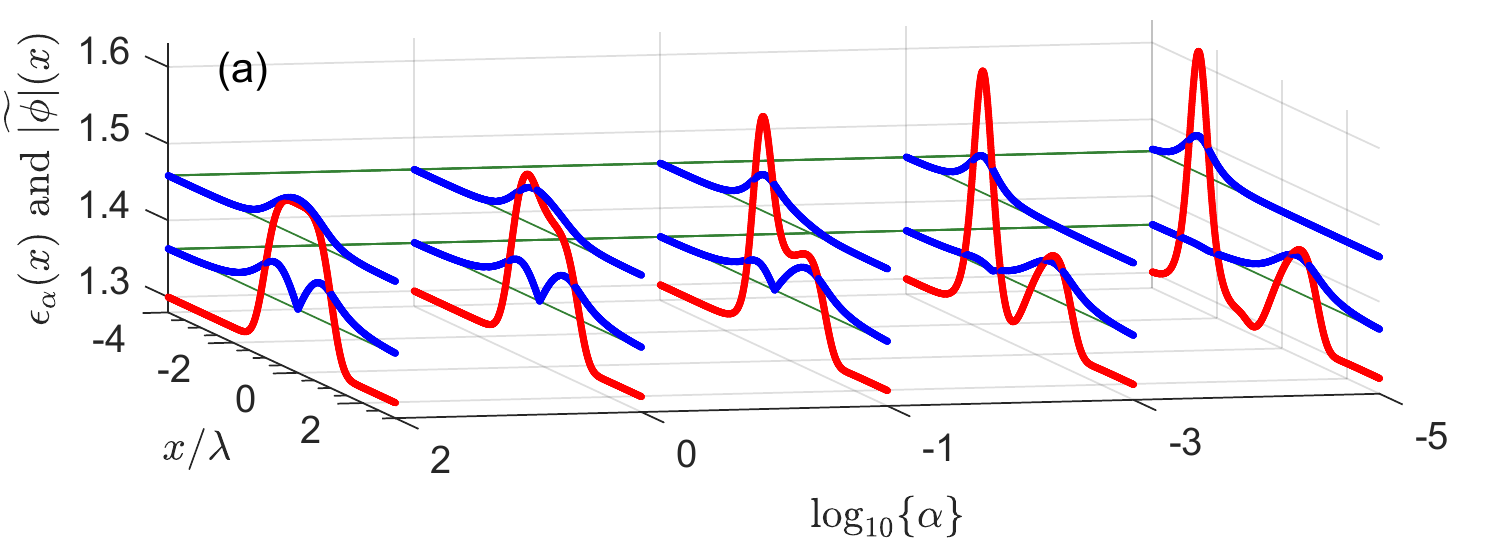}
	\includegraphics[width = \columnwidth, clip = true, trim = {20 5 55 200}]{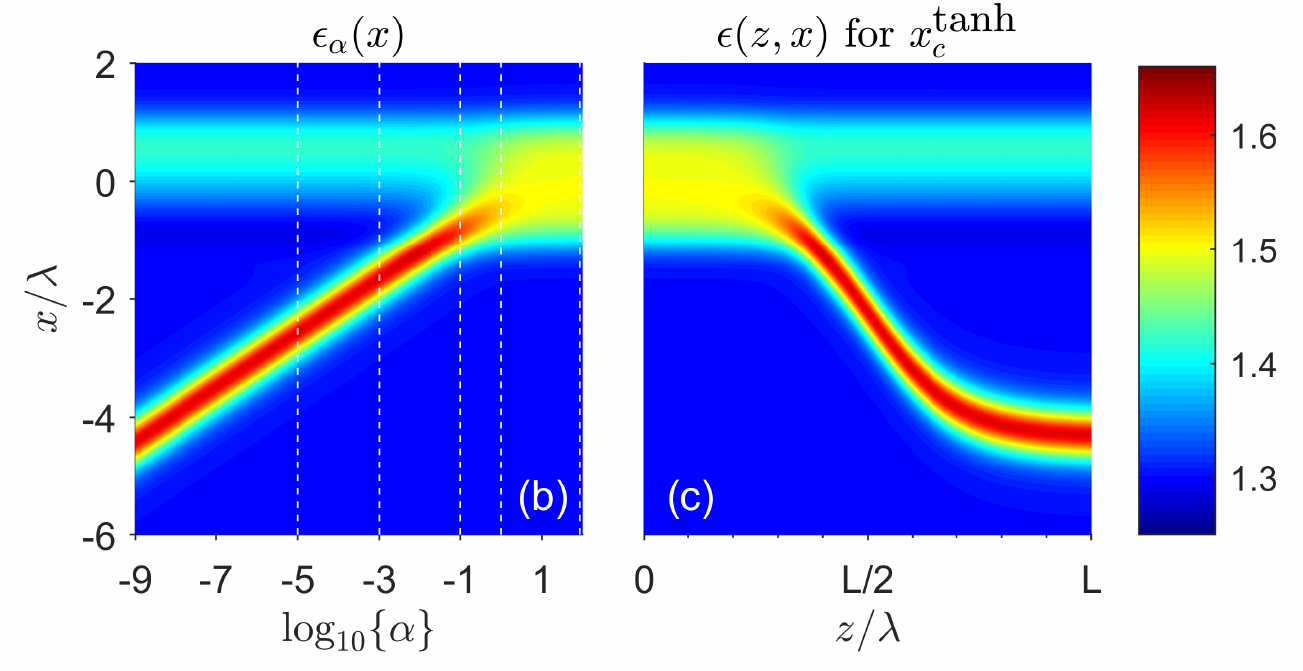}
	\caption{
		(a),~Permittivity profiles $\epsilon_{\alpha}(x)$ (red lines) for the values of the parameter $\alpha$ marked by white dashed lines in~(b), for $\epsilon_{\infty}(x) = 1.3+\Delta \epsilon \exp[(x/\sigma)^4]$. 
		The normalized mode amplitudes $\widetilde{|\phi_j|}(x)$ of a given permittivity profile are shown with blue lines~\cite{w1}. 
		The eigenvalues $n_{\rm{eff},j}^2$ for each of the modes are marked by the green lines.
		The parameters used are: $\Delta\epsilon = 0.2$, $\sigma = 1.1 \lambda$.
		(b)~Landscape of the permittivity $\epsilon_{\alpha}(x)$ created using the SUSY transformation given by~\cref{eqn:susy}. 
		(c)~Permittivity landscape for the SUSY-based mode sorter for which the center of the waveguide supporting the fundamental mode follows the trajectory $x_{c}^{\rm{tanh}}(z) = \frac{H}{2} [\tanh(A\xi)+1]$, for $A=5$, $H = -4.2\lambda$, and $\xi = \frac{z}{L}-\frac{1}{2}$.
	}
	\label{fig:profiles}
\end{figure}

\Cref{eqn:Helm} has the form of a time-independent Schr\"odinger equation and therefore, the permittivity distributions iso-spectral with $\epsilon_{\infty}$ can be found using a continuous SUSY transformation.
The single-parameter family of iso-spectral permittivity distributions is described by~\cite{Cooper01}:
\begin{equation}
\epsilon_{\alpha}(x) = \epsilon_{\infty}(x) + 2k_0^{-2} \partial_x^2 \ln[I_1(x)+\alpha],
\label{eqn:susy}
\end{equation}
where the parameter $\alpha \in \mathbb{R} \backslash [-1,0]$, and $I_1(x) = \int_{-\infty}^{x} \phi_1^2(x') \textrm{d}x'$. 
Here, $\phi_1$ denotes the fundamental TE-polarized mode of $\epsilon_{\infty}$.
Permittivity profiles of a few members of the iso-spectral family~$\epsilon_{\alpha}$ are shown in \cref{fig:profiles}(a) with the supported mode profiles~$\phi$.
\Cref{fig:profiles}(b) shows the permittivity distribution $\epsilon_{\alpha}(x)$ as a function of the parameter $\alpha$.
For $\alpha\rightarrow\pm\infty$, one recovers the original permittivity profile.
Decreasing the value of $|\alpha|$ modifies the shape of the original potential in such a way that for $\alpha \rightarrow 0$ ($-1$) a waveguide supporting the fundamental mode emerges and shifts towards $x= -\infty$ ($+\infty$).

Using the continuous SUSY transformation outlined above, we design a mode sorter for two modes in the following steps.
We start with the input (multimode) waveguide profile given by $\epsilon_{\infty}(x)$.
For this waveguide, we perform the SUSY transformation described by \cref{eqn:susy}, which results in a family of permittivity profiles $\epsilon_{\alpha}$ shown in \cref{fig:profiles}(b).
From there, we can extract the relation between the position of the center of the waveguide supporting the fundamental mode $x_{c}$ and the parameter $\alpha$. 
Now, for a mode sorter with a geometry given by an arbitrary $x_{c}(z)$, we can find the appropriate mapping of $\alpha(z)$.
As a result, we obtain a permittivty landscape $\epsilon(x,z)$ like the one shown in \cref{fig:profiles}(c), where a SUSY mode sorter described by $x_{c}^{\rm{tanh}}(z) = \frac{H}{2} [\tanh(A\xi)+1]$ is shown. Here, $\xi = \frac{z}{L}-\frac{1}{2}$, and $L$ denotes the length of the sorter.
The value of the parameter $H$ is chosen in such a way that the overlap between the modes at the output of the sorter is smaller than 0.1\% 
In a sorter designed using this method, the input modes are spatially separated, and, at the same time, the higher order modes of the input waveguide are converted into the fundamental modes of the output waveguides while the propagation constants of all the modes are preserved along the propagation direction.

\section{Results}

\subsection{Methods}

In order to estimate the performance of the mode sorter designed using the continuous SUSY transformation, we have used two methods. 
In the first one, we excite separately each of the modes supported by the input waveguide and study the cross-talk to the output waveguides that do not support the excited mode.
The cross-talk $C_{i,j}$ is defined as the power in the output waveguide supporting the mode $j$ divided by the input power while only the mode $i$ is excited.
In this method, the optimal performance of the coupler is achieved by minimizing the cross-talk and maximizing the power in the output waveguide supporting the excited mode (minimizing the losses of the sorter).

\begin{figure}[!t]
	\includegraphics[width = 0.99\columnwidth, clip = true, trim = {20 185 35 5}]{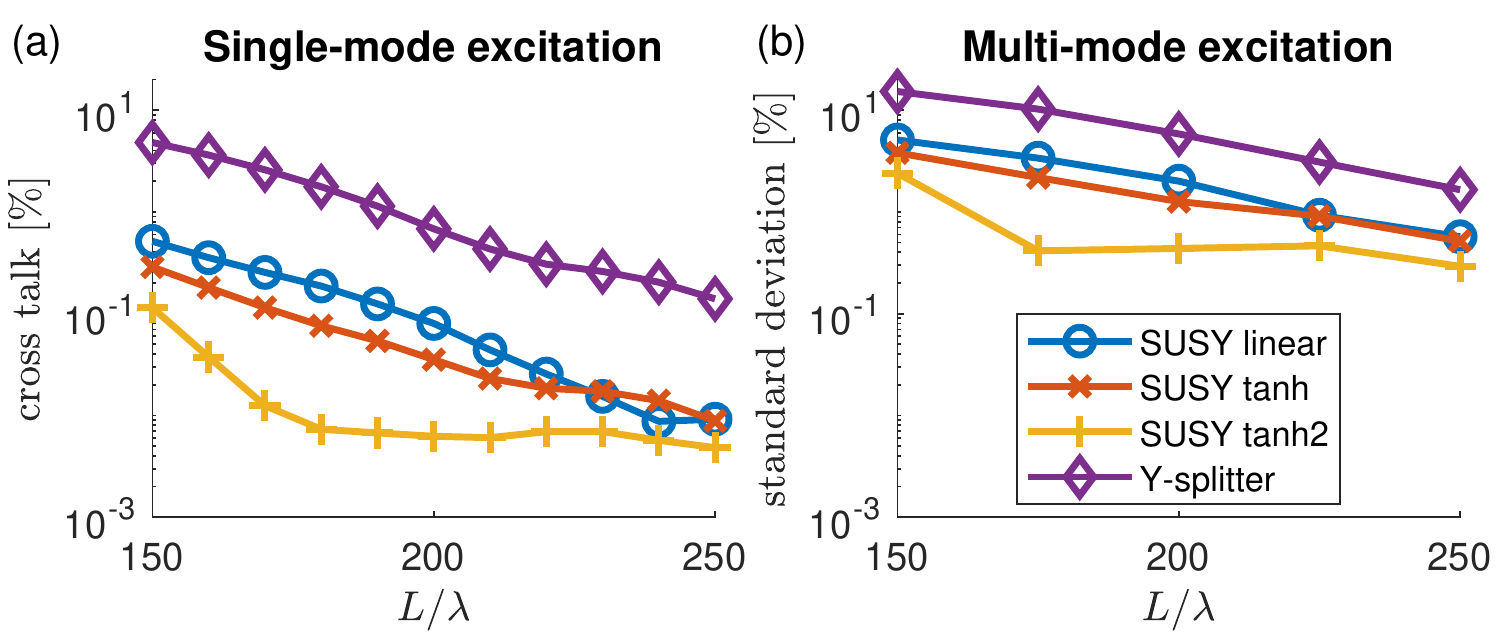}\\
	\includegraphics[width = 0.477\columnwidth, clip = true, trim = {45 0 365 0}]{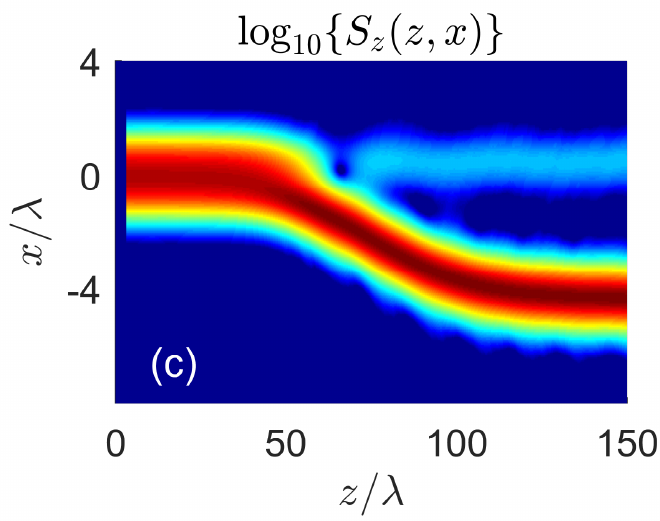}
	\includegraphics[width = 0.503\columnwidth, clip = true, trim = {335 0 65 0}]{Fig22.pdf}
	\caption{ 
		Performance comparison of a standard asymmetric Y-splitter and the three SUSY mode sorters described in the text using two methods: single-mode excitation (a), and multi-mode excitation (b).
		(a)~Cross-talk ($\max\{C_{1,2}, C_{2,1}\}$) as a function of the sorter length $L$.
		(b)~Standard deviation of the transmittance averaged over the input phase difference between the modes ($\max\{\rm{std}[T_1], \rm{std}[T_2]\}$) as a function of the sorter length $L$.
		(c),~(d) Typical distribution of the $z$-component of the normalized Poynting vector $\mathbf{S}$ for single-mode excitation (fundamental mode excited) (c) and multi-mode excitation (d).
	}
	\label{fig:shape}
\end{figure}

In the second method, all the modes of the input waveguide are excited at the same time, in such a way that they inject equal powers into the sorter. 
Here, we define the transmittance $T_i$ as the ratio between the power in the output waveguide supporting the mode $i$ and the total input power. 
Because of the interference between the excited modes, the transmittances $T_i$ depend on the phase difference between the input modes. 
We compute the transmittances for the values of the phase difference between $0$ to $2\pi$ and compute the standard deviation of the obtained results.
Here, the optimum performance is obtained when all the $T_i$'s are equal to each other and their standard deviations are minimized.

\subsection{Sorter geometry}

\Cref{fig:shape} shows the analysis of the SUSY mode-sorter efficiency using the two methods outlined above for three different geometries $x_c(z)$. 
We compare their performance with a standard asymmetric Y-splitter used as a reference. 
For the SUSY sorters, the position of the center of the waveguide supporting the fundamental mode is given by (i) a linear function $x_c^{\rm{linear}}(z) = \frac{H}{L}z$, (ii) $x_c^{\rm{tanh}}(z)$, and (iii) $x_{c}^{\rm{tanh2}}(z) = \frac{H}{2}\{\tanh  B\frac{\xi}{R}   + C \,\mathrm{sign}(\xi) (\frac{\xi}{R})^m ]+1\}$. 
Here, $R = L/(L-2 z_0)$, where $z_0$ is the solution of $\partial_z x_{c}^{\rm{tanh2}}(z)=d_0$, which ensures that the inclination of the waveguide center profile at the input and output is smaller that $d_0=10^{-3}$.
The parameters $B=5$, $C=60$, and $m=4$ were found using an optimization procedure aimed at minimizing the mode cross-talk for a fixed sorter length $L$. 

For the asymmetric Y-splitter used as a reference, we have chosen a rectangular input waveguide described by $\epsilon_{Y}(x) = 1.3+\Delta \epsilon_Y \mathcal{H}(-|x|+\sigma_Y)$, where $\mathcal{H}$ is the Heaviside step function, $\sigma_Y = 1.1\lambda$, and $\Delta \epsilon_Y=0.2$. 
This input waveguide splits into two asymmetric rectangular branches with the widths $\sigma_1$ and $\sigma_2$ ($\sigma_1 + \sigma_2 = \sigma_Y$). 
The separation between these two branches is then linearly increased along the length of the splitter.
The width of the two branches were optimized to yield the lowest cross-talk for a given length, resulting in $\sigma_1=0.4\sigma_Y$.
All the simulations of the light propagation in the mode sorter were performed using the finite-difference time-domain solver implemented in Lumerical~\cite{lumerical}.


Figures~\ref{fig:shape}(a),~\ref{fig:shape}(b) reveal that the single- and  multi-mode excitation methods yield the same dependence of the sorter performance both on the length and on geometry, as expected in a linear system. 
The modal cross-talk in the sorters described by $x_c^{\rm{linear}}$ and $x_c^{\rm{tanh}}$, as well as the Y-splitter decreases linearly with the increase of the length $L$.
For the geometry given by $x_c^{\rm{tanh2}}$, the cross-talk decreases rapidly for $L<175\lambda$ and then remains below the level of 0.01\% ($-40$~dB).
However, despite the low values of the cross-talk, the standard deviation of the power in the two arms has much larger values as a result of the interference pattern forming in the input waveguide. 
As the phase difference at the input changes, the position of the interference maximum periodically shifts from one edge of the input waveguide to the other (along the $x$-direction). 
Depending on the spatial location of the interference maximum, the power coupled to the output branches of the sorter varies. 
For longer devices the standard deviation of the power is minimized, as the coupling region becomes longer and contains multiple periods of the interference pattern. 
Typical distributions of the $z$-component of the Poynting vector obtained using the two methods are shown in Figs.~\ref{fig:shape}(c) and \ref{fig:shape}(d). 
As both methods give qualitatively the same results, in the following, we will only use the single-mode-excitation method since it is less computationally demanding (no averaging over the input phase difference required).

The comparison of the performance in terms of the cross-talk of the SUSY-based linear sorter with a standard asymmetric Y-splitter shows that the SUSY-based sorter outperforms the Y-splitter by at least one order of magnitude. 
Using a smooth $x_c^{\rm{tanh}}(z)$ profile allows for further reduction of the cross-talk. 
Another order of magnitude improvement can be achieved by further optimizing the $x_c(z)$ profile and using the sorter described by $x_c^{\rm{tanh2}}$. 
For all the SUSY-sorter geometries $x_c$ and lengths $L$ studied in \cref{fig:shape}, the power lost in the mode sorter remains below~1\%.

\begin{figure}[!b]
	\includegraphics[width = \columnwidth, clip = true, trim = {10 5 30 15}]{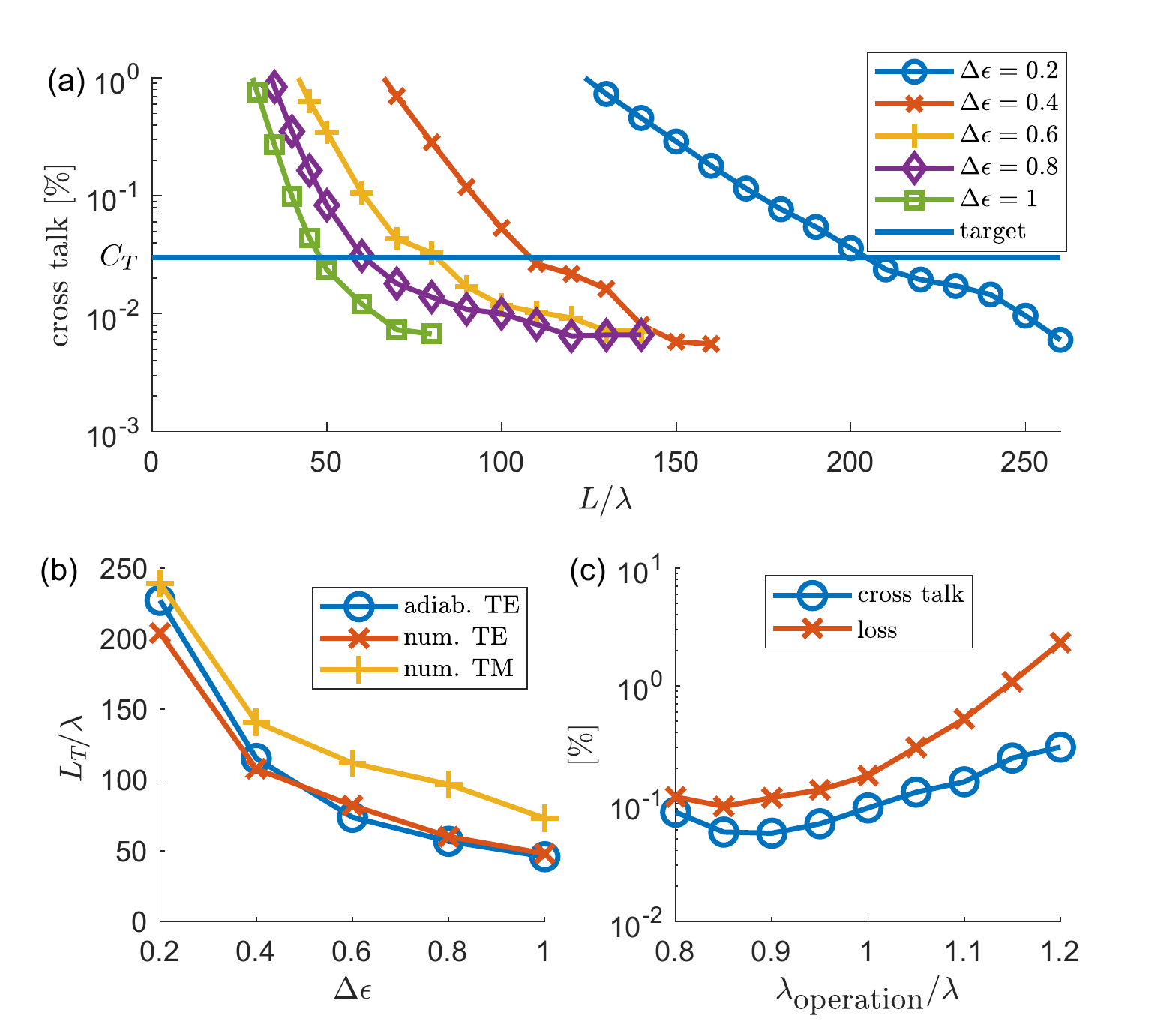}
	\caption{
		(a) Cross-talk in the SUSY mode sorter described by the $x_c^{\tanh}$ as a function of the sorter length $L$ for different permittivity contrasts $\Delta \epsilon$.
		With the increase of the permittivity contrast the width of the initial permittivity profile $\epsilon_{\infty}$ is reduced ($\sigma = [1.1, 0.8, 0.6, 0.55, 0.5]\lambda$) so that only two input modes are supported. 
		(b)~Comparison between the sorter length at which the cross-talk reaches the target value ($C_T=0.03\%$, shown by blue line) ($\times$) and the length for which the coupling probability predicted using the adiabaticity condition is below 0.01\% ($\circ$).
		The $+$ symbols show the performance of the mode sorter for the TM input polarization obtained using the numerical simulations.
		(c)~Cross-talk and the power lost in the mode sorter as a function of the operation wavelength for $L=200\lambda$, $\Delta \epsilon = 0.2$, and $\sigma=0.9\lambda$.
	}
	\label{fig:broadTETM}
\end{figure}

\subsection{Permittivity-contrast, wavelength, and polarization sensitivity}

\Cref{fig:broadTETM}(a) shows the dependence of the performance of the mode sorter described by $x_c^{\rm{tanh}}$ on the length~$L$ for different values of the permittivity contrast~$\Delta \epsilon$. 
It can be seen that increasing the permittivity contrast allows to reduce the device length necessary to obtain the cross-talk below a certain target value.
Moreover, the length at which the target cross-talk is reached can be predicted from the adiabacity condition~\cite{Menchon16}. 
The probability of excitation of the mode $j$ while the energy is concentrated in the mode $i$ is defined as $p_{i \rightarrow j} \le \max\left| \left< \phi_j(z) \left|\partial_z\right| \phi_i(z) \right> |\beta_i(z)-\beta_j(z)|^{-1} \right|^2$. 
\Cref{fig:broadTETM}(b) shows the comparison between the sorter lengths $L_T$ at which the target cross-talk is reached ($C_T = 3\cdot10^{-3}$) in the numerical simulations and the lengths for which the coupling probability predicted using the adiabacity condition is smaller than $10^{-3}$.
The results obtained using these two methods are in good agreement.

Despite the fact that the mode sorter is designed using the Eqs.~(\ref{eqn:Helm}) and (\ref{eqn:susy}) for the TE light polarization, it can efficiently operate for the TM polarization. 
As seen in \cref{fig:broadTETM}(b), the performance for the TM polarization of the input light is slightly decreased compared to the TE excitation.
Nevertheless, the cross-talk value below $C_T$ can be reached if the device length is increased by about~1.5 times.

The SUSY design procedure described by \cref{eqn:susy} uses the fundamental mode of the input waveguide that is computed at a fixed design wavelength $\lambda$. 
In spite of that, as shown in \cref{fig:broadTETM}(c), the SUSY mode sorter can operate in a broad range of wavelengths.
The cross-talk remains at the level below 0.4\% while the losses are maintained below 2.5\% in a wavelength range spanning 40\% of the design wavelength. 
In the wavelength range under consideration, the input waveguide supports precisely two modes.
The performance of the sorter improves at shorter operation wavelengths due to the two following reasons.
Firstly, the device becomes longer with respect to the operation wavelength and therefore the adiabacity improves.
Secondly, the separation between the two output branches increases with respect to the operation wavelength and, 
as a result, the coupling between the branches decreases.

\begin{figure}[!b]
	\includegraphics[width = 0.59\columnwidth, clip = true, trim = {0 0 25 0}]{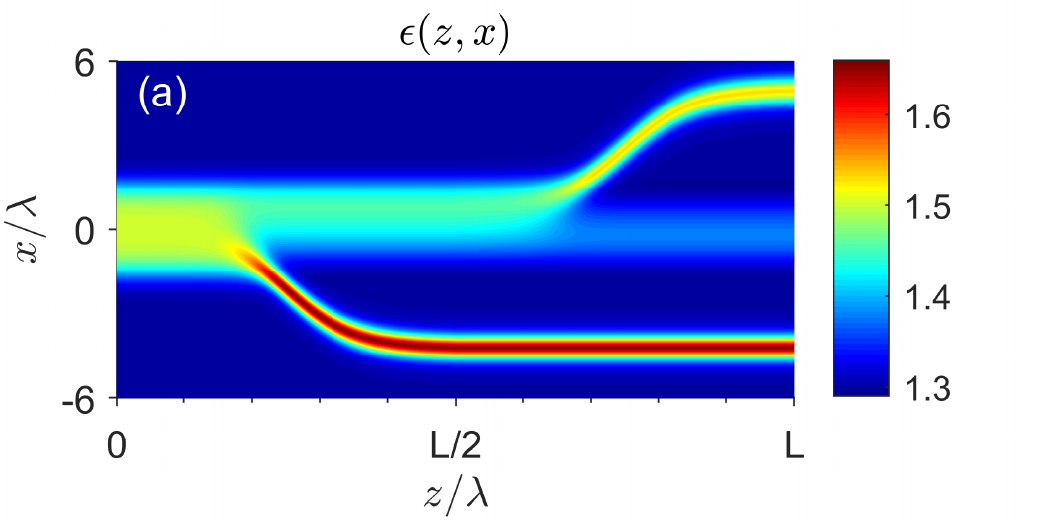}
	\includegraphics[width = 0.39\columnwidth, clip = true, trim = {0 0 5 0}]{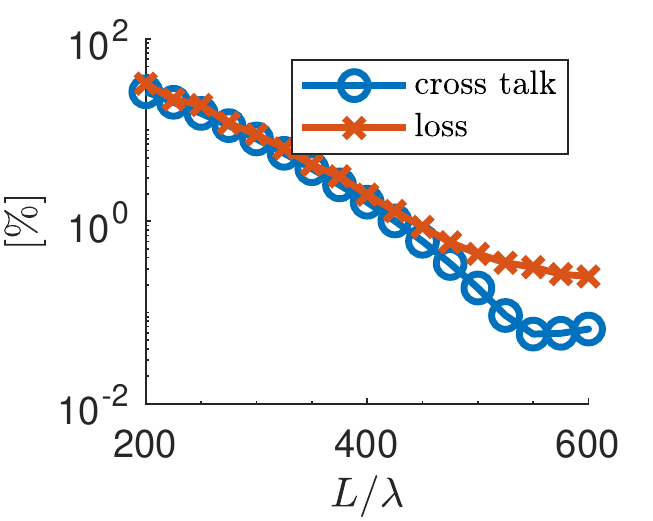}
	\caption{
		(a)~Permittivity landscape for the SUSY sorter capable of spatially separating three modes.
		(b)~Dependence of the cross-talk ($\max\{C_{1,2}, C_{1,3}, C_{2,1}, C_{2,3}, C_{3,1}, C_{3,2}\}$) and power lost in the sorter as a function of the sorter length $L$.
	}
	\label{fig:3modes}
\end{figure}

\subsection{Sorting multiple modes}

Finally we show that it is possible to design a SUSY-based mode sorter capable of separating multiple modes. 
As illustrated in \cref{fig:3modes}(a) the modes are sorted in a sequential manner.
In the first step, the fundamental mode of the input waveguide is separated. 
In the second step, the second order mode of the input waveguide becomes the fundamental mode of the top waveguide (located around $x=0$) at $z=L/2$. 
Therefore, this new fundamental mode can be used in the SUSY transformation described by~\cref{eqn:susy} applied to  the permittivity profile $\epsilon(x,z=L/2)$. 
This time we use $\alpha<-1$ in order to shift the center of the newly emerged waveguide in the positive $x$-direction.
The resulting permittivity landscape is shown in \cref{fig:3modes}(a). 
As it can be seen in \cref{fig:3modes}(b), for a sufficiently long sorter ($L>500\lambda$) the mode cross-talk and the device losses remain below 0.1\% and 1\%, respectively.

\section{Summary}

In conclusion, we have presented a design of a mode sorter constructed using iso-spectral permittivity profiles generated by the continuous transformation in the broken supersymmetric regime.
In this design, the propagation constants of the modes are preserved over the entire length of the sorter which, in connection with the adiabatic permittivity modification along the propagation direction, allows to minimize the cross-talk between the output waveguides and the scattering losses. 
As a result, we report two orders of magnitude reduction of the cross-talk compared to a standard asymmetric Y-splitters.
Even though the supersymmetric mode sorters are designed at a specific wavelength and for the transverse-electric light polarization, their performance is not compromised in a broad wavelength range spanning 40\% of the design wavelength and for transverse-magnetic polarization.
Finally, we have demonstrated that the supersymmetry based design can be used to efficiently sort multiple modes. 
The experimental demonstration of the proposed modes sorter might be enabled by tapered optical fibers~\cite{Black88}, channel waveguide segmentation~\cite{Weissman92,Chou:96}, femtosecond laser written techniques~\cite{DellaValle09}, or electron-beam lithography~\cite{CHEN15,JIANG17}.
The supersymmetry-based optimization method of the mode converter presented here can be used as an alternative for other design techniques based on adiabatic mode evolution. 
For instance, the mode sorters designed using the procedure described in this work achieve comparable cross-talk and device length as the state-of-the-art systems designed using the method based on fast quasiadiabatic dynamics~\cite{Martinez-Garaot:17,Chung:17}.

\section*{Acknowledgments}

This work was supported by Army Research Office (ARO) (W911NF-17-1-0400)

\bibliography{Walasik-SUSY-mode-sorter}

\begin{thebibliography}{44}%
\makeatletter
\providecommand \@ifxundefined [1]{%
 \@ifx{#1\undefined}
}%
\providecommand \@ifnum [1]{%
 \ifnum #1\expandafter \@firstoftwo
 \else \expandafter \@secondoftwo
 \fi
}%
\providecommand \@ifx [1]{%
 \ifx #1\expandafter \@firstoftwo
 \else \expandafter \@secondoftwo
 \fi
}%
\providecommand \natexlab [1]{#1}%
\providecommand \enquote  [1]{``#1''}%
\providecommand \bibnamefont  [1]{#1}%
\providecommand \bibfnamefont [1]{#1}%
\providecommand \citenamefont [1]{#1}%
\providecommand \href@noop [0]{\@secondoftwo}%
\providecommand \href [0]{\begingroup \@sanitize@url \@href}%
\providecommand \@href[1]{\@@startlink{#1}\@@href}%
\providecommand \@@href[1]{\endgroup#1\@@endlink}%
\providecommand \@sanitize@url [0]{\catcode `\\12\catcode `\$12\catcode
  `\&12\catcode `\#12\catcode `\^12\catcode `\_12\catcode `\%12\relax}%
\providecommand \@@startlink[1]{}%
\providecommand \@@endlink[0]{}%
\providecommand \url  [0]{\begingroup\@sanitize@url \@url }%
\providecommand \@url [1]{\endgroup\@href {#1}{\urlprefix }}%
\providecommand \urlprefix  [0]{URL }%
\providecommand \Eprint [0]{\href }%
\providecommand \doibase [0]{http://dx.doi.org/}%
\providecommand \selectlanguage [0]{\@gobble}%
\providecommand \bibinfo  [0]{\@secondoftwo}%
\providecommand \bibfield  [0]{\@secondoftwo}%
\providecommand \translation [1]{[#1]}%
\providecommand \BibitemOpen [0]{}%
\providecommand \bibitemStop [0]{}%
\providecommand \bibitemNoStop [0]{.\EOS\space}%
\providecommand \EOS [0]{\spacefactor3000\relax}%
\providecommand \BibitemShut  [1]{\csname bibitem#1\endcsname}%
\let\auto@bib@innerbib\@empty
\bibitem [{\citenamefont {Agrell}\ \emph {et~al.}(2016)\citenamefont {Agrell},
  \citenamefont {Karlsson}, \citenamefont {Chraplyvy}, \citenamefont
  {Richardson}, \citenamefont {Krummrich}, \citenamefont {Winzer},
  \citenamefont {Roberts}, \citenamefont {Fischer}, \citenamefont {Savory},
  \citenamefont {Eggleton}, \citenamefont {Secondini}, \citenamefont
  {Kschischang}, \citenamefont {Lord}, \citenamefont {Prat}, \citenamefont
  {Tomkos}, \citenamefont {Bowers}, \citenamefont {Srinivasan}, \citenamefont
  {Brandt-Pearce},\ and\ \citenamefont {Gisin}}]{Agrell16}%
  \BibitemOpen
  \bibfield  {author} {\bibinfo {author} {\bibfnamefont {E.}~\bibnamefont
  {Agrell}}, \bibinfo {author} {\bibfnamefont {M.}~\bibnamefont {Karlsson}},
  \bibinfo {author} {\bibfnamefont {A.~R.}\ \bibnamefont {Chraplyvy}}, \bibinfo
  {author} {\bibfnamefont {D.~J.}\ \bibnamefont {Richardson}}, \bibinfo
  {author} {\bibfnamefont {P.~M.}\ \bibnamefont {Krummrich}}, \bibinfo {author}
  {\bibfnamefont {P.}~\bibnamefont {Winzer}}, \bibinfo {author} {\bibfnamefont
  {K.}~\bibnamefont {Roberts}}, \bibinfo {author} {\bibfnamefont {J.~K.}\
  \bibnamefont {Fischer}}, \bibinfo {author} {\bibfnamefont {S.~J.}\
  \bibnamefont {Savory}}, \bibinfo {author} {\bibfnamefont {B.~J.}\
  \bibnamefont {Eggleton}}, \bibinfo {author} {\bibfnamefont {M.}~\bibnamefont
  {Secondini}}, \bibinfo {author} {\bibfnamefont {F.~R.}\ \bibnamefont
  {Kschischang}}, \bibinfo {author} {\bibfnamefont {A.}~\bibnamefont {Lord}},
  \bibinfo {author} {\bibfnamefont {J.}~\bibnamefont {Prat}}, \bibinfo {author}
  {\bibfnamefont {I.}~\bibnamefont {Tomkos}}, \bibinfo {author} {\bibfnamefont
  {J.~E.}\ \bibnamefont {Bowers}}, \bibinfo {author} {\bibfnamefont
  {S.}~\bibnamefont {Srinivasan}}, \bibinfo {author} {\bibfnamefont
  {M.}~\bibnamefont {Brandt-Pearce}}, \ and\ \bibinfo {author} {\bibfnamefont
  {N.}~\bibnamefont {Gisin}},\ }\bibfield  {title} {\enquote {\bibinfo {title}
  {Roadmap of optical communications},}\ }\href
  {http://stacks.iop.org/2040-8986/18/i=6/a=063002} {\bibfield  {journal}
  {\bibinfo  {journal} {J. Opt.}\ }\textbf {\bibinfo {volume} {18}},\ \bibinfo
  {pages} {063002} (\bibinfo {year} {2016})}\BibitemShut {NoStop}%
\bibitem [{\citenamefont {Lifante}(2003)}]{Lifante03}%
  \BibitemOpen
  \bibfield  {author} {\bibinfo {author} {\bibfnamefont {G.}~\bibnamefont
  {Lifante}},\ }\href@noop {} {\emph {\bibinfo {title} {Integrated Photonics:
  Fundamentals}}}\ (\bibinfo  {publisher} {John Willey \& Sons, Ltd},\ \bibinfo
  {year} {2003})\BibitemShut {NoStop}%
\bibitem [{\citenamefont {Reed}\ and\ \citenamefont {Knights}(2004)}]{Reed04}%
  \BibitemOpen
  \bibfield  {author} {\bibinfo {author} {\bibfnamefont {G.~T.}\ \bibnamefont
  {Reed}}\ and\ \bibinfo {author} {\bibfnamefont {A.~P.}\ \bibnamefont
  {Knights}},\ }\href@noop {} {\emph {\bibinfo {title} {Silicon Photonics, An
  Introduction}}}\ (\bibinfo  {publisher} {John Willey \& Sons, Ltd},\ \bibinfo
  {year} {2004})\BibitemShut {NoStop}%
\bibitem [{\citenamefont {Burns}\ and\ \citenamefont {Milton}(1974)}]{Burns74}%
  \BibitemOpen
  \bibfield  {author} {\bibinfo {author} {\bibfnamefont {U.~K.}\ \bibnamefont
  {Burns}}\ and\ \bibinfo {author} {\bibfnamefont {A.~F.}\ \bibnamefont
  {Milton}},\ }\bibfield  {title} {\enquote {\bibinfo {title} {Mode conversion
  in planar dielectric separating waveguides},}\ }in\ \href {\doibase
  10.1109/IEDM.1974.6219637} {\emph {\bibinfo {booktitle} {1974 International
  Electron Devices Meeting (IEDM)}}}\ (\bibinfo {year} {1974})\ pp.\ \bibinfo
  {pages} {81--82}\BibitemShut {NoStop}%
\bibitem [{\citenamefont {Love}\ and\ \citenamefont {Riesen}(2012)}]{Love12}%
  \BibitemOpen
  \bibfield  {author} {\bibinfo {author} {\bibfnamefont {J.~D.}\ \bibnamefont
  {Love}}\ and\ \bibinfo {author} {\bibfnamefont {N.}~\bibnamefont {Riesen}},\
  }\bibfield  {title} {\enquote {\bibinfo {title} {Single-, few-, and multimode
  y-junctions},}\ }\href {\doibase 10.1109/JLT.2011.2179976} {\bibfield
  {journal} {\bibinfo  {journal} {J. Light. Technol.}\ }\textbf {\bibinfo
  {volume} {30}},\ \bibinfo {pages} {304--309} (\bibinfo {year}
  {2012})}\BibitemShut {NoStop}%
\bibitem [{\citenamefont {Riesen}\ and\ \citenamefont
  {Love}(2012)}]{Riesen:12}%
  \BibitemOpen
  \bibfield  {author} {\bibinfo {author} {\bibfnamefont {N.}~\bibnamefont
  {Riesen}}\ and\ \bibinfo {author} {\bibfnamefont {J.~D.}\ \bibnamefont
  {Love}},\ }\bibfield  {title} {\enquote {\bibinfo {title} {Design of
  mode-sorting asymmetric y-junctions},}\ }\href {\doibase
  10.1364/AO.51.002778} {\bibfield  {journal} {\bibinfo  {journal} {Appl.
  Opt.}\ }\textbf {\bibinfo {volume} {51}},\ \bibinfo {pages} {2778--2783}
  (\bibinfo {year} {2012})}\BibitemShut {NoStop}%
\bibitem [{\citenamefont {Driscoll}\ \emph {et~al.}(2013)\citenamefont
  {Driscoll}, \citenamefont {Grote}, \citenamefont {Souhan}, \citenamefont
  {Dadap}, \citenamefont {Lu},\ and\ \citenamefont {Osgood}}]{Driscoll:13}%
  \BibitemOpen
  \bibfield  {author} {\bibinfo {author} {\bibfnamefont {J.~B.}\ \bibnamefont
  {Driscoll}}, \bibinfo {author} {\bibfnamefont {R.~R.}\ \bibnamefont {Grote}},
  \bibinfo {author} {\bibfnamefont {B.}~\bibnamefont {Souhan}}, \bibinfo
  {author} {\bibfnamefont {J.~I.}\ \bibnamefont {Dadap}}, \bibinfo {author}
  {\bibfnamefont {M.}~\bibnamefont {Lu}}, \ and\ \bibinfo {author}
  {\bibfnamefont {R.~M.}\ \bibnamefont {Osgood}},\ }\bibfield  {title}
  {\enquote {\bibinfo {title} {Asymmetric y junctions in silicon waveguides for
  on-chip mode-division multiplexing},}\ }\href {\doibase 10.1364/OL.38.001854}
  {\bibfield  {journal} {\bibinfo  {journal} {Opt. Lett.}\ }\textbf {\bibinfo
  {volume} {38}},\ \bibinfo {pages} {1854--1856} (\bibinfo {year}
  {2013})}\BibitemShut {NoStop}%
\bibitem [{\citenamefont {Zhang}\ \emph {et~al.}(2013)\citenamefont {Zhang},
  \citenamefont {Yang}, \citenamefont {Lim}, \citenamefont {Lo}, \citenamefont
  {Galland}, \citenamefont {Baehr-Jones},\ and\ \citenamefont
  {Hochberg}}]{Zhang:13}%
  \BibitemOpen
  \bibfield  {author} {\bibinfo {author} {\bibfnamefont {Y.}~\bibnamefont
  {Zhang}}, \bibinfo {author} {\bibfnamefont {S.}~\bibnamefont {Yang}},
  \bibinfo {author} {\bibfnamefont {A.~E.-J.}\ \bibnamefont {Lim}}, \bibinfo
  {author} {\bibfnamefont {G.-Q.}\ \bibnamefont {Lo}}, \bibinfo {author}
  {\bibfnamefont {C.}~\bibnamefont {Galland}}, \bibinfo {author} {\bibfnamefont
  {T.}~\bibnamefont {Baehr-Jones}}, \ and\ \bibinfo {author} {\bibfnamefont
  {M.}~\bibnamefont {Hochberg}},\ }\bibfield  {title} {\enquote {\bibinfo
  {title} {A compact and low loss y-junction for submicron silicon
  waveguide},}\ }\href {\doibase 10.1364/OE.21.001310} {\bibfield  {journal}
  {\bibinfo  {journal} {Opt. Express}\ }\textbf {\bibinfo {volume} {21}},\
  \bibinfo {pages} {1310--1316} (\bibinfo {year} {2013})}\BibitemShut {NoStop}%
\bibitem [{\citenamefont {Birks}\ \emph {et~al.}(2015)\citenamefont {Birks},
  \citenamefont {Gris-S\'{a}nchez}, \citenamefont {Yerolatsitis}, \citenamefont
  {Leon-Saval},\ and\ \citenamefont {Thomson}}]{Birks:15}%
  \BibitemOpen
  \bibfield  {author} {\bibinfo {author} {\bibfnamefont {T.~A.}\ \bibnamefont
  {Birks}}, \bibinfo {author} {\bibfnamefont {I.}~\bibnamefont
  {Gris-S\'{a}nchez}}, \bibinfo {author} {\bibfnamefont {S.}~\bibnamefont
  {Yerolatsitis}}, \bibinfo {author} {\bibfnamefont {S.~G.}\ \bibnamefont
  {Leon-Saval}}, \ and\ \bibinfo {author} {\bibfnamefont {R.~R.}\ \bibnamefont
  {Thomson}},\ }\bibfield  {title} {\enquote {\bibinfo {title} {The photonic
  lantern},}\ }\href {\doibase 10.1364/AOP.7.000107} {\bibfield  {journal}
  {\bibinfo  {journal} {Adv. Opt. Photon.}\ }\textbf {\bibinfo {volume} {7}},\
  \bibinfo {pages} {107--167} (\bibinfo {year} {2015})}\BibitemShut {NoStop}%
\bibitem [{\citenamefont {Dai}(2017)}]{Dai17}%
  \BibitemOpen
  \bibfield  {author} {\bibinfo {author} {\bibfnamefont {D.}~\bibnamefont
  {Dai}},\ }\bibfield  {title} {\enquote {\bibinfo {title} {Silicon
  nanophotonic integrated devices for on-chip multiplexing and switching},}\
  }\href {\doibase 10.1109/JLT.2016.2587727} {\bibfield  {journal} {\bibinfo
  {journal} {J. Light. Technol.}\ }\textbf {\bibinfo {volume} {35}},\ \bibinfo
  {pages} {572--587} (\bibinfo {year} {2017})}\BibitemShut {NoStop}%
\bibitem [{\citenamefont {Fischer}\ \emph {et~al.}(1994)\citenamefont
  {Fischer}, \citenamefont {Zinke}, \citenamefont {Schuppert},\ and\
  \citenamefont {Petermann}}]{Fischer94}%
  \BibitemOpen
  \bibfield  {author} {\bibinfo {author} {\bibfnamefont {U.}~\bibnamefont
  {Fischer}}, \bibinfo {author} {\bibfnamefont {T.}~\bibnamefont {Zinke}},
  \bibinfo {author} {\bibfnamefont {B.}~\bibnamefont {Schuppert}}, \ and\
  \bibinfo {author} {\bibfnamefont {K.}~\bibnamefont {Petermann}},\ }\bibfield
  {title} {\enquote {\bibinfo {title} {Singlemode optical switches based on soi
  waveguides with large cross-section},}\ }\href {\doibase 10.1049/el:19940310}
  {\bibfield  {journal} {\bibinfo  {journal} {Electron. Lett.}\ }\textbf
  {\bibinfo {volume} {30}},\ \bibinfo {pages} {406--408} (\bibinfo {year}
  {1994})}\BibitemShut {NoStop}%
\bibitem [{\citenamefont {Sun}\ \emph {et~al.}(2009)\citenamefont {Sun},
  \citenamefont {Liu},\ and\ \citenamefont {Yariv}}]{Sun:09}%
  \BibitemOpen
  \bibfield  {author} {\bibinfo {author} {\bibfnamefont {X.}~\bibnamefont
  {Sun}}, \bibinfo {author} {\bibfnamefont {H.-C.}\ \bibnamefont {Liu}}, \ and\
  \bibinfo {author} {\bibfnamefont {A.}~\bibnamefont {Yariv}},\ }\bibfield
  {title} {\enquote {\bibinfo {title} {Adiabaticity criterion and the shortest
  adiabatic mode transformer in a coupled-waveguide system},}\ }\href {\doibase
  10.1364/OL.34.000280} {\bibfield  {journal} {\bibinfo  {journal} {Opt.
  Lett.}\ }\textbf {\bibinfo {volume} {34}},\ \bibinfo {pages} {280--282}
  (\bibinfo {year} {2009})}\BibitemShut {NoStop}%
\bibitem [{\citenamefont {Mart\'{i}nez-Garaot}\ \emph
  {et~al.}(2014)\citenamefont {Mart\'{i}nez-Garaot}, \citenamefont {Tseng},\
  and\ \citenamefont {Muga}}]{Martinez-Garaot:14}%
  \BibitemOpen
  \bibfield  {author} {\bibinfo {author} {\bibfnamefont {S.}~\bibnamefont
  {Mart\'{i}nez-Garaot}}, \bibinfo {author} {\bibfnamefont {S.-Y.}\
  \bibnamefont {Tseng}}, \ and\ \bibinfo {author} {\bibfnamefont {J.~G.}\
  \bibnamefont {Muga}},\ }\bibfield  {title} {\enquote {\bibinfo {title}
  {Compact and high conversion efficiency mode-sorting asymmetric y junction
  using shortcuts to adiabaticity},}\ }\href {\doibase 10.1364/OL.39.002306}
  {\bibfield  {journal} {\bibinfo  {journal} {Opt. Lett.}\ }\textbf {\bibinfo
  {volume} {39}},\ \bibinfo {pages} {2306--2309} (\bibinfo {year}
  {2014})}\BibitemShut {NoStop}%
\bibitem [{\citenamefont {Mart\'{i}nez-Garaot}\ \emph
  {et~al.}(2017)\citenamefont {Mart\'{i}nez-Garaot}, \citenamefont {Muga},\
  and\ \citenamefont {Tseng}}]{Martinez-Garaot:17}%
  \BibitemOpen
  \bibfield  {author} {\bibinfo {author} {\bibfnamefont {S.}~\bibnamefont
  {Mart\'{i}nez-Garaot}}, \bibinfo {author} {\bibfnamefont {J.~G.}\
  \bibnamefont {Muga}}, \ and\ \bibinfo {author} {\bibfnamefont {S.-Y.}\
  \bibnamefont {Tseng}},\ }\bibfield  {title} {\enquote {\bibinfo {title}
  {Shortcuts to adiabaticity in optical waveguides using fast quasiadiabatic
  dynamics},}\ }\href {\doibase 10.1364/OE.25.000159} {\bibfield  {journal}
  {\bibinfo  {journal} {Opt. Express}\ }\textbf {\bibinfo {volume} {25}},\
  \bibinfo {pages} {159--167} (\bibinfo {year} {2017})}\BibitemShut {NoStop}%
\bibitem [{\citenamefont {Chung}\ \emph {et~al.}(2017)\citenamefont {Chung},
  \citenamefont {Lee},\ and\ \citenamefont {Tseng}}]{Chung:17}%
  \BibitemOpen
  \bibfield  {author} {\bibinfo {author} {\bibfnamefont {H.-C.}\ \bibnamefont
  {Chung}}, \bibinfo {author} {\bibfnamefont {K.-S.}\ \bibnamefont {Lee}}, \
  and\ \bibinfo {author} {\bibfnamefont {S.-Y.}\ \bibnamefont {Tseng}},\
  }\bibfield  {title} {\enquote {\bibinfo {title} {Short and broadband silicon
  asymmetric y-junction two-mode (de)multiplexer using fast quasiadiabatic
  dynamics},}\ }\href {\doibase 10.1364/OE.25.013626} {\bibfield  {journal}
  {\bibinfo  {journal} {Opt. Express}\ }\textbf {\bibinfo {volume} {25}},\
  \bibinfo {pages} {13626--13634} (\bibinfo {year} {2017})}\BibitemShut
  {NoStop}%
\bibitem [{\citenamefont {Heinrich}\ \emph {et~al.}(2014)\citenamefont
  {Heinrich}, \citenamefont {Miri}, \citenamefont {St\"{u}tzer}, \citenamefont
  {El-Ganainy}, \citenamefont {Nolte}, \citenamefont {Szameit},\ and\
  \citenamefont {Christodoulides}}]{Heinrich:14a}%
  \BibitemOpen
  \bibfield  {author} {\bibinfo {author} {\bibfnamefont {M.}~\bibnamefont
  {Heinrich}}, \bibinfo {author} {\bibfnamefont {M.-A.}\ \bibnamefont {Miri}},
  \bibinfo {author} {\bibfnamefont {S.}~\bibnamefont {St\"{u}tzer}}, \bibinfo
  {author} {\bibfnamefont {R.}~\bibnamefont {El-Ganainy}}, \bibinfo {author}
  {\bibfnamefont {S.}~\bibnamefont {Nolte}}, \bibinfo {author} {\bibfnamefont
  {A.}~\bibnamefont {Szameit}}, \ and\ \bibinfo {author} {\bibfnamefont
  {D.~N.}\ \bibnamefont {Christodoulides}},\ }\bibfield  {title} {\enquote
  {\bibinfo {title} {Supersymmetric mode converters},}\ }\href {\doibase
  10.1038/ncomms4698} {\bibfield  {journal} {\bibinfo  {journal} {Nat.
  Commun.}\ }\textbf {\bibinfo {volume} {6}},\ \bibinfo {pages} {3698}
  (\bibinfo {year} {2014})}\BibitemShut {NoStop}%
\bibitem [{\citenamefont {Queralt\'{o}}\ \emph {et~al.}(2017)\citenamefont
  {Queralt\'{o}}, \citenamefont {Ahufinger},\ and\ \citenamefont
  {Mompart}}]{Queralto:17}%
  \BibitemOpen
  \bibfield  {author} {\bibinfo {author} {\bibfnamefont {G.}~\bibnamefont
  {Queralt\'{o}}}, \bibinfo {author} {\bibfnamefont {V.}~\bibnamefont
  {Ahufinger}}, \ and\ \bibinfo {author} {\bibfnamefont {J.}~\bibnamefont
  {Mompart}},\ }\bibfield  {title} {\enquote {\bibinfo {title} {Mode-division
  (de)multiplexing using adiabatic passage and supersymmetric waveguides},}\
  }\href {\doibase 10.1364/OE.25.027396} {\bibfield  {journal} {\bibinfo
  {journal} {Opt. Express}\ }\textbf {\bibinfo {volume} {25}},\ \bibinfo
  {pages} {27396--27404} (\bibinfo {year} {2017})}\BibitemShut {NoStop}%
\bibitem [{\citenamefont {Queralt\'{o}}\ \emph {et~al.}(2018)\citenamefont
  {Queralt\'{o}}, \citenamefont {Ahufinger},\ and\ \citenamefont
  {Mompart}}]{Queralto:18}%
  \BibitemOpen
  \bibfield  {author} {\bibinfo {author} {\bibfnamefont {G.}~\bibnamefont
  {Queralt\'{o}}}, \bibinfo {author} {\bibfnamefont {V.}~\bibnamefont
  {Ahufinger}}, \ and\ \bibinfo {author} {\bibfnamefont {J.}~\bibnamefont
  {Mompart}},\ }\bibfield  {title} {\enquote {\bibinfo {title} {Integrated
  photonic devices based on adiabatic transitions between supersymmetric
  structures},}\ }\href {\doibase 10.1364/OE.26.033797} {\bibfield  {journal}
  {\bibinfo  {journal} {Opt. Express}\ }\textbf {\bibinfo {volume} {26}},\
  \bibinfo {pages} {33797--33806} (\bibinfo {year} {2018})}\BibitemShut
  {NoStop}%
\bibitem [{\citenamefont {Weinberg}(2005)}]{Weinberg05}%
  \BibitemOpen
  \bibfield  {author} {\bibinfo {author} {\bibfnamefont {S.}~\bibnamefont
  {Weinberg}},\ }\href@noop {} {\emph {\bibinfo {title} {The Quantum Theory of
  Fields, Supersymmetry}}}\ (\bibinfo  {publisher} {Cambridge University
  Press},\ \bibinfo {year} {2005})\BibitemShut {NoStop}%
\bibitem [{\citenamefont {Cooper}\ \emph {et~al.}(2001)\citenamefont {Cooper},
  \citenamefont {Khare},\ and\ \citenamefont {Sukhatme}}]{Cooper01}%
  \BibitemOpen
  \bibfield  {author} {\bibinfo {author} {\bibfnamefont {F.}~\bibnamefont
  {Cooper}}, \bibinfo {author} {\bibfnamefont {A.}~\bibnamefont {Khare}}, \
  and\ \bibinfo {author} {\bibfnamefont {U.}~\bibnamefont {Sukhatme}},\
  }\href@noop {} {\emph {\bibinfo {title} {Supersymmetry in Quantum
  Mechanics}}}\ (\bibinfo  {publisher} {World Scientific},\ \bibinfo {year}
  {2001})\BibitemShut {NoStop}%
\bibitem [{\citenamefont {Maydanyuk}(2005)}]{MAYDANYUK05}%
  \BibitemOpen
  \bibfield  {author} {\bibinfo {author} {\bibfnamefont {S.~P.}\ \bibnamefont
  {Maydanyuk}},\ }\bibfield  {title} {\enquote {\bibinfo {title}
  {Susy-hierarchy of one-dimensional reflectionless potentials},}\ }\href
  {\doibase https://doi.org/10.1016/j.aop.2004.11.004} {\bibfield  {journal}
  {\bibinfo  {journal} {Ann. Phys.}\ }\textbf {\bibinfo {volume} {316}},\
  \bibinfo {pages} {440 -- 465} (\bibinfo {year} {2005})}\BibitemShut {NoStop}%
\bibitem [{\citenamefont {Dunne}\ and\ \citenamefont
  {Feinberg}(1998)}]{Dunne98a}%
  \BibitemOpen
  \bibfield  {author} {\bibinfo {author} {\bibfnamefont {G.}~\bibnamefont
  {Dunne}}\ and\ \bibinfo {author} {\bibfnamefont {J.}~\bibnamefont
  {Feinberg}},\ }\bibfield  {title} {\enquote {\bibinfo {title}
  {Self-isospectral periodic potentials and supersymmetric quantum
  mechanics},}\ }\href {\doibase 10.1103/PhysRevD.57.1271} {\bibfield
  {journal} {\bibinfo  {journal} {Phys. Rev. D}\ }\textbf {\bibinfo {volume}
  {57}},\ \bibinfo {pages} {1271--1276} (\bibinfo {year} {1998})}\BibitemShut
  {NoStop}%
\bibitem [{\citenamefont {Khare}\ and\ \citenamefont
  {Sukhatme}(2004)}]{Khare04}%
  \BibitemOpen
  \bibfield  {author} {\bibinfo {author} {\bibfnamefont {A.}~\bibnamefont
  {Khare}}\ and\ \bibinfo {author} {\bibfnamefont {U.}~\bibnamefont
  {Sukhatme}},\ }\bibfield  {title} {\enquote {\bibinfo {title} {Periodic
  potentials and supersymmetry},}\ }\href
  {http://stacks.iop.org/0305-4470/37/i=43/a=002} {\bibfield  {journal}
  {\bibinfo  {journal} {J. Phys. A}\ }\textbf {\bibinfo {volume} {37}},\
  \bibinfo {pages} {10037} (\bibinfo {year} {2004})}\BibitemShut {NoStop}%
\bibitem [{\citenamefont {Chumakov}\ and\ \citenamefont
  {Wolf}(1994)}]{CHUMAKOV94}%
  \BibitemOpen
  \bibfield  {author} {\bibinfo {author} {\bibfnamefont {S.~M.}\ \bibnamefont
  {Chumakov}}\ and\ \bibinfo {author} {\bibfnamefont {K.~B.}\ \bibnamefont
  {Wolf}},\ }\bibfield  {title} {\enquote {\bibinfo {title} {Supersymmetry in
  helmholtz optics},}\ }\href {\doibase
  https://doi.org/10.1016/0375-9601(94)00616-4} {\bibfield  {journal} {\bibinfo
   {journal} {Phys.Lett. A}\ }\textbf {\bibinfo {volume} {193}},\ \bibinfo
  {pages} {51 -- 53} (\bibinfo {year} {1994})}\BibitemShut {NoStop}%
\bibitem [{\citenamefont {Miri}\ \emph {et~al.}(2013)\citenamefont {Miri},
  \citenamefont {Heinrich}, \citenamefont {El-Ganainy},\ and\ \citenamefont
  {Christodoulides}}]{Miri13}%
  \BibitemOpen
  \bibfield  {author} {\bibinfo {author} {\bibfnamefont {M.-A.}\ \bibnamefont
  {Miri}}, \bibinfo {author} {\bibfnamefont {M.}~\bibnamefont {Heinrich}},
  \bibinfo {author} {\bibfnamefont {R.}~\bibnamefont {El-Ganainy}}, \ and\
  \bibinfo {author} {\bibfnamefont {D.~N.}\ \bibnamefont {Christodoulides}},\
  }\bibfield  {title} {\enquote {\bibinfo {title} {Supersymmetric optical
  structures},}\ }\href {\doibase 10.1103/PhysRevLett.110.233902} {\bibfield
  {journal} {\bibinfo  {journal} {Phys. Rev. Lett.}\ }\textbf {\bibinfo
  {volume} {110}},\ \bibinfo {pages} {233902} (\bibinfo {year}
  {2013})}\BibitemShut {NoStop}%
\bibitem [{\citenamefont {Miri}\ \emph {et~al.}(2014)\citenamefont {Miri},
  \citenamefont {Heinrich},\ and\ \citenamefont {Christodoulides}}]{Miri:14}%
  \BibitemOpen
  \bibfield  {author} {\bibinfo {author} {\bibfnamefont {M.-A.}\ \bibnamefont
  {Miri}}, \bibinfo {author} {\bibfnamefont {M.}~\bibnamefont {Heinrich}}, \
  and\ \bibinfo {author} {\bibfnamefont {D.~N.}\ \bibnamefont
  {Christodoulides}},\ }\bibfield  {title} {\enquote {\bibinfo {title}
  {Susy-inspired one-dimensional transformation optics},}\ }\href {\doibase
  10.1364/OPTICA.1.000089} {\bibfield  {journal} {\bibinfo  {journal} {Optica}\
  }\textbf {\bibinfo {volume} {1}},\ \bibinfo {pages} {89--95} (\bibinfo {year}
  {2014})}\BibitemShut {NoStop}%
\bibitem [{\citenamefont {Laba}\ and\ \citenamefont {Tkachuk}(2014)}]{Laba14}%
  \BibitemOpen
  \bibfield  {author} {\bibinfo {author} {\bibfnamefont {H.~P.}\ \bibnamefont
  {Laba}}\ and\ \bibinfo {author} {\bibfnamefont {V.~M.}\ \bibnamefont
  {Tkachuk}},\ }\bibfield  {title} {\enquote {\bibinfo {title}
  {Quantum-mechanical analogy and supersymmetry of electromagnetic wave modes
  in planar waveguides},}\ }\href {\doibase 10.1103/PhysRevA.89.033826}
  {\bibfield  {journal} {\bibinfo  {journal} {Phys. Rev. A}\ }\textbf {\bibinfo
  {volume} {89}},\ \bibinfo {pages} {033826} (\bibinfo {year}
  {2014})}\BibitemShut {NoStop}%
\bibitem [{\citenamefont {Midya}\ \emph {et~al.}(2018)\citenamefont {Midya},
  \citenamefont {Walasik}, \citenamefont {Litchinitser},\ and\ \citenamefont
  {Feng}}]{Midya:18}%
  \BibitemOpen
  \bibfield  {author} {\bibinfo {author} {\bibfnamefont {B.}~\bibnamefont
  {Midya}}, \bibinfo {author} {\bibfnamefont {W.}~\bibnamefont {Walasik}},
  \bibinfo {author} {\bibfnamefont {N.~M.}\ \bibnamefont {Litchinitser}}, \
  and\ \bibinfo {author} {\bibfnamefont {L.}~\bibnamefont {Feng}},\ }\bibfield
  {title} {\enquote {\bibinfo {title} {Supercharge optical arrays},}\ }\href
  {\doibase 10.1364/OL.43.004927} {\bibfield  {journal} {\bibinfo  {journal}
  {Opt. Lett.}\ }\textbf {\bibinfo {volume} {43}},\ \bibinfo {pages}
  {4927--4930} (\bibinfo {year} {2018})}\BibitemShut {NoStop}%
\bibitem [{\citenamefont {Milanovi\ifmmode~\acute{c}\else \'{c}\fi{}}\ and\
  \citenamefont {Ikoni\ifmmode~\acute{c}\else \'{c}\fi{}}(1996)}]{Milanovic96}%
  \BibitemOpen
  \bibfield  {author} {\bibinfo {author} {\bibfnamefont {V.}~\bibnamefont
  {Milanovi\ifmmode~\acute{c}\else \'{c}\fi{}}}\ and\ \bibinfo {author}
  {\bibfnamefont {Z.}~\bibnamefont {Ikoni\ifmmode~\acute{c}\else \'{c}\fi{}}},\
  }\bibfield  {title} {\enquote {\bibinfo {title} {On the optimization of
  resonant intersubband nonlinear optical susceptibilities in semiconductor
  quantum wells},}\ }\href {\doibase 10.1109/3.511544} {\bibfield  {journal}
  {\bibinfo  {journal} {IEEE J. Quantum Electron.}\ }\textbf {\bibinfo {volume}
  {32}},\ \bibinfo {pages} {1316--1323} (\bibinfo {year} {1996})}\BibitemShut
  {NoStop}%
\bibitem [{\citenamefont {Tomi\ifmmode~\acute{c}\else \'{c}\fi{}}\ \emph
  {et~al.}(1997)\citenamefont {Tomi\ifmmode~\acute{c}\else \'{c}\fi{}},
  \citenamefont {Milanovi\ifmmode~\acute{c}\else \'{c}\fi{}},\ and\
  \citenamefont {Ikoni\ifmmode~\acute{c}\else \'{c}\fi{}}}]{Tomic97}%
  \BibitemOpen
  \bibfield  {author} {\bibinfo {author} {\bibfnamefont {S.}~\bibnamefont
  {Tomi\ifmmode~\acute{c}\else \'{c}\fi{}}}, \bibinfo {author} {\bibfnamefont
  {V.}~\bibnamefont {Milanovi\ifmmode~\acute{c}\else \'{c}\fi{}}}, \ and\
  \bibinfo {author} {\bibfnamefont {Z.}~\bibnamefont
  {Ikoni\ifmmode~\acute{c}\else \'{c}\fi{}}},\ }\bibfield  {title} {\enquote
  {\bibinfo {title} {Optimization of intersubband resonant second-order
  susceptibility in asymmetric graded
  al${}_{x}{\mathrm{ga}}_{1\ensuremath{-}x}$as quantum wells using
  supersymmetric quantum mechanics},}\ }\href {\doibase
  10.1103/PhysRevB.56.1033} {\bibfield  {journal} {\bibinfo  {journal} {Phys.
  Rev. B}\ }\textbf {\bibinfo {volume} {56}},\ \bibinfo {pages} {1033--1036}
  (\bibinfo {year} {1997})}\BibitemShut {NoStop}%
\bibitem [{\citenamefont {Bai}\ and\ \citenamefont {Citrin}(2008)}]{Bai:08}%
  \BibitemOpen
  \bibfield  {author} {\bibinfo {author} {\bibfnamefont {J.}~\bibnamefont
  {Bai}}\ and\ \bibinfo {author} {\bibfnamefont {D.~S.}\ \bibnamefont
  {Citrin}},\ }\bibfield  {title} {\enquote {\bibinfo {title} {Enhancement of
  optical kerr effect in quantum-cascade lasers with multiple resonance
  levels},}\ }\href {\doibase 10.1364/OE.16.012599} {\bibfield  {journal}
  {\bibinfo  {journal} {Opt. Express}\ }\textbf {\bibinfo {volume} {16}},\
  \bibinfo {pages} {12599--12606} (\bibinfo {year} {2008})}\BibitemShut
  {NoStop}%
\bibitem [{\citenamefont {El-Ganainy}\ \emph {et~al.}(2015)\citenamefont
  {El-Ganainy}, \citenamefont {Ge}, \citenamefont {Khajavikhan},\ and\
  \citenamefont {Christodoulides}}]{El-Ganainy15}%
  \BibitemOpen
  \bibfield  {author} {\bibinfo {author} {\bibfnamefont {R.}~\bibnamefont
  {El-Ganainy}}, \bibinfo {author} {\bibfnamefont {Li}~\bibnamefont {Ge}},
  \bibinfo {author} {\bibfnamefont {M.}~\bibnamefont {Khajavikhan}}, \ and\
  \bibinfo {author} {\bibfnamefont {D.~N.}\ \bibnamefont {Christodoulides}},\
  }\bibfield  {title} {\enquote {\bibinfo {title} {Supersymmetric laser
  arrays},}\ }\href {\doibase 10.1103/PhysRevA.92.033818} {\bibfield  {journal}
  {\bibinfo  {journal} {Phys. Rev. A}\ }\textbf {\bibinfo {volume} {92}},\
  \bibinfo {pages} {033818} (\bibinfo {year} {2015})}\BibitemShut {NoStop}%
\bibitem [{\citenamefont {Teimourpour}\ \emph {et~al.}(2016)\citenamefont
  {Teimourpour}, \citenamefont {Ge},\ and\ \citenamefont
  {El-Ganainy}}]{Teimourpour16}%
  \BibitemOpen
  \bibfield  {author} {\bibinfo {author} {\bibfnamefont {M.~H.}\ \bibnamefont
  {Teimourpour}}, \bibinfo {author} {\bibfnamefont {L.}~\bibnamefont {Ge}}, \
  and\ \bibinfo {author} {\bibfnamefont {R.}~\bibnamefont {El-Ganainy}},\
  }\bibfield  {title} {\enquote {\bibinfo {title} {Non-hermitian engineering of
  single mode two dimensional laser arrays},}\ }\href {\doibase
  10.1038/srep33253} {\bibfield  {journal} {\bibinfo  {journal} {Sci. Rep.}\
  }\textbf {\bibinfo {volume} {6}},\ \bibinfo {pages} {33253} (\bibinfo {year}
  {2016})}\BibitemShut {NoStop}%
\bibitem [{\citenamefont {Hokmabadi}\ \emph {et~al.}(2018)\citenamefont
  {Hokmabadi}, \citenamefont {Nye}, \citenamefont {El-Ganainy}, \citenamefont
  {Christodoulides},\ and\ \citenamefont {Khajavikhan}}]{Hokmabadi2:18}%
  \BibitemOpen
  \bibfield  {author} {\bibinfo {author} {\bibfnamefont {M.~P.}\ \bibnamefont
  {Hokmabadi}}, \bibinfo {author} {\bibfnamefont {N.~S.}\ \bibnamefont {Nye}},
  \bibinfo {author} {\bibfnamefont {R.}~\bibnamefont {El-Ganainy}}, \bibinfo
  {author} {\bibfnamefont {D.~N.}\ \bibnamefont {Christodoulides}}, \ and\
  \bibinfo {author} {\bibfnamefont {M.}~\bibnamefont {Khajavikhan}},\
  }\bibfield  {title} {\enquote {\bibinfo {title} {Supersymmetric laser
  arrays},}\ }\href {https://arxiv.org/abs/1812.10690} {\bibfield  {journal}
  {\bibinfo  {journal} {arXiv:1812{.}10690}\ } (\bibinfo {year}
  {2018})}\BibitemShut {NoStop}%
\bibitem [{\citenamefont {Walasik}\ \emph {et~al.}(2018)\citenamefont
  {Walasik}, \citenamefont {Midya}, \citenamefont {Feng},\ and\ \citenamefont
  {Litchinitser}}]{Walasik:18}%
  \BibitemOpen
  \bibfield  {author} {\bibinfo {author} {\bibfnamefont {W.}~\bibnamefont
  {Walasik}}, \bibinfo {author} {\bibfnamefont {B.}~\bibnamefont {Midya}},
  \bibinfo {author} {\bibfnamefont {L.}~\bibnamefont {Feng}}, \ and\ \bibinfo
  {author} {\bibfnamefont {N.~M.}\ \bibnamefont {Litchinitser}},\ }\bibfield
  {title} {\enquote {\bibinfo {title} {Supersymmetry-guided method for mode
  selection and optimization in coupled systems},}\ }\href {\doibase
  10.1364/OL.43.003758} {\bibfield  {journal} {\bibinfo  {journal} {Opt.
  Lett.}\ }\textbf {\bibinfo {volume} {43}},\ \bibinfo {pages} {3758--3761}
  (\bibinfo {year} {2018})}\BibitemShut {NoStop}%
\bibitem [{\citenamefont {Menchon-Enrich}\ \emph {et~al.}(2016)\citenamefont
  {Menchon-Enrich}, \citenamefont {Benseny}, \citenamefont {Ahufinger},
  \citenamefont {Greentree}, \citenamefont {Busch},\ and\ \citenamefont
  {Mompart}}]{Menchon16}%
  \BibitemOpen
  \bibfield  {author} {\bibinfo {author} {\bibfnamefont {R.}~\bibnamefont
  {Menchon-Enrich}}, \bibinfo {author} {\bibfnamefont {A.}~\bibnamefont
  {Benseny}}, \bibinfo {author} {\bibfnamefont {V.}~\bibnamefont {Ahufinger}},
  \bibinfo {author} {\bibfnamefont {A.~D.}\ \bibnamefont {Greentree}}, \bibinfo
  {author} {\bibfnamefont {Th.}\ \bibnamefont {Busch}}, \ and\ \bibinfo
  {author} {\bibfnamefont {J.}~\bibnamefont {Mompart}},\ }\bibfield  {title}
  {\enquote {\bibinfo {title} {Spatial adiabatic passage: a review of recent
  progress},}\ }\href {http://stacks.iop.org/0034-4885/79/i=7/a=074401}
  {\bibfield  {journal} {\bibinfo  {journal} {Rep. Prog. Phys.}\ }\textbf
  {\bibinfo {volume} {79}},\ \bibinfo {pages} {074401} (\bibinfo {year}
  {2016})}\BibitemShut {NoStop}%
\bibitem [{w1()}]{w1}%
  \BibitemOpen
  \href@noop {} {}\bibinfo {note} {The normalized modes amplitudes are plotted
  according to the expression $\widetilde{|\phi_j|} = 0.5 (n_{\rm{eff},1}^2 -
  n_{\rm{eff},2}^2) |\phi_j|/ \max[|\phi_j|]+n_{\rm{eff},j}^2$.}\BibitemShut
  {Stop}%
\bibitem [{\citenamefont {{Lumerical Inc.}}(2019)}]{lumerical}%
  \BibitemOpen
  \bibfield  {author} {\bibinfo {author} {\bibnamefont {{Lumerical Inc.}}},\
  }\href@noop {} {\enquote {\bibinfo {title}
  {http://www.lumerical.com/tcad-products/fdtd/},}\ } (\bibinfo {year}
  {2019})\BibitemShut {NoStop}%
\bibitem [{\citenamefont {Black}\ \emph {et~al.}(1988)\citenamefont {Black},
  \citenamefont {Gonthier}, \citenamefont {Lacroix}, \citenamefont {Lapierre},\
  and\ \citenamefont {Bures}}]{Black88}%
  \BibitemOpen
  \bibfield  {author} {\bibinfo {author} {\bibfnamefont {R.~J.}\ \bibnamefont
  {Black}}, \bibinfo {author} {\bibfnamefont {E}~\bibnamefont {Gonthier}},
  \bibinfo {author} {\bibfnamefont {S.}~\bibnamefont {Lacroix}}, \bibinfo
  {author} {\bibfnamefont {J.}~\bibnamefont {Lapierre}}, \ and\ \bibinfo
  {author} {\bibfnamefont {J.}~\bibnamefont {Bures}},\ }\bibfield  {title}
  {\enquote {\bibinfo {title} {Tapered fibers: An overview},}\ }in\ \href
  {\doibase 10.1117/12.942540} {\emph {\bibinfo {booktitle} {Proc. SPIE 0839,
  Components for Fiber Optic Applications II}}}\ (\bibinfo {year} {1988})\
  p.~\bibinfo {pages} {18}\BibitemShut {NoStop}%
\bibitem [{\citenamefont {Weissman}\ and\ \citenamefont
  {Hardy}(1992)}]{Weissman92}%
  \BibitemOpen
  \bibfield  {author} {\bibinfo {author} {\bibfnamefont {Z.}~\bibnamefont
  {Weissman}}\ and\ \bibinfo {author} {\bibfnamefont {A.}~\bibnamefont
  {Hardy}},\ }\bibfield  {title} {\enquote {\bibinfo {title} {2-d mode tapering
  via tapered channel waveguide segmentation},}\ }\href {\doibase
  10.1049/el:19920962} {\bibfield  {journal} {\bibinfo  {journal} {Electron.
  Lett.}\ }\textbf {\bibinfo {volume} {28}},\ \bibinfo {pages} {1514--1516}
  (\bibinfo {year} {1992})}\BibitemShut {NoStop}%
\bibitem [{\citenamefont {Chou}\ \emph {et~al.}(1996)\citenamefont {Chou},
  \citenamefont {Arbore},\ and\ \citenamefont {Fejer}}]{Chou:96}%
  \BibitemOpen
  \bibfield  {author} {\bibinfo {author} {\bibfnamefont {M.~H.}\ \bibnamefont
  {Chou}}, \bibinfo {author} {\bibfnamefont {M.~A.}\ \bibnamefont {Arbore}}, \
  and\ \bibinfo {author} {\bibfnamefont {M.~M.}\ \bibnamefont {Fejer}},\
  }\bibfield  {title} {\enquote {\bibinfo {title} {Adiabatically tapered
  periodic segmentation of channel waveguides for mode-size transformation and
  fundamental mode excitation},}\ }\href {\doibase 10.1364/OL.21.000794}
  {\bibfield  {journal} {\bibinfo  {journal} {Opt. Lett.}\ }\textbf {\bibinfo
  {volume} {21}},\ \bibinfo {pages} {794--796} (\bibinfo {year}
  {1996})}\BibitemShut {NoStop}%
\bibitem [{\citenamefont {Valle}\ \emph {et~al.}(2009)\citenamefont {Valle},
  \citenamefont {Osellame},\ and\ \citenamefont {Laporta}}]{DellaValle09}%
  \BibitemOpen
  \bibfield  {author} {\bibinfo {author} {\bibfnamefont {G.~Della}\
  \bibnamefont {Valle}}, \bibinfo {author} {\bibfnamefont {R.}~\bibnamefont
  {Osellame}}, \ and\ \bibinfo {author} {\bibfnamefont {P.}~\bibnamefont
  {Laporta}},\ }\bibfield  {title} {\enquote {\bibinfo {title} {Micromachining
  of photonic devices by femtosecond laser pulses},}\ }\href
  {http://stacks.iop.org/1464-4258/11/i=1/a=013001} {\bibfield  {journal}
  {\bibinfo  {journal} {J. Opt. A: Pure Appl. Opt}\ }\textbf {\bibinfo {volume}
  {11}},\ \bibinfo {pages} {013001} (\bibinfo {year} {2009})}\BibitemShut
  {NoStop}%
\bibitem [{\citenamefont {Chen}(2015)}]{CHEN15}%
  \BibitemOpen
  \bibfield  {author} {\bibinfo {author} {\bibfnamefont {Y.}~\bibnamefont
  {Chen}},\ }\bibfield  {title} {\enquote {\bibinfo {title} {Nanofabrication by
  electron beam lithography and its applications: A review},}\ }\href {\doibase
  https://doi.org/10.1016/j.mee.2015.02.042} {\bibfield  {journal} {\bibinfo
  {journal} {Microelectron. Eng.}\ }\textbf {\bibinfo {volume} {135}},\
  \bibinfo {pages} {57 -- 72} (\bibinfo {year} {2015})}\BibitemShut {NoStop}%
\bibitem [{\citenamefont {Jiang}(2017)}]{JIANG17}%
  \BibitemOpen
  \bibfield  {author} {\bibinfo {author} {\bibfnamefont {N.}~\bibnamefont
  {Jiang}},\ }\bibfield  {title} {\enquote {\bibinfo {title} {On the spatial
  resolution limit of direct-write electron beam lithography},}\ }\href
  {\doibase https://doi.org/10.1016/j.mee.2016.10.016} {\bibfield  {journal}
  {\bibinfo  {journal} {Microelectron. Eng.}\ }\textbf {\bibinfo {volume}
  {168}},\ \bibinfo {pages} {41 -- 44} (\bibinfo {year} {2017})}\BibitemShut
  {NoStop}%
\end{thebibliography}%

\end{document}